# Bound Whispering Gallery Modes in Circular Arrays of Dielectric Spherical Particles

Alexander L. Burin[*a], Gail S. Blaustein,[b] Olga M. Samoilova[a]

[a] Department of Chemistry and [b] Department of Mathematics, Tulane University, New Orleans, LA, USA 70118-5636

**ABSTRACT**

Low-dimensional ordered arrays of dielectric particles can possess bound optical modes having an extremely high quality factor depending on the material used. If these arrays consist of metal particles, then they cannot have a high quality factor because their light absorption restricts performance. In this paper we address the following question: can bound modes be formed in dielectric systems where the absorption of light is negligible? Our investigation of circular arrays of spherical particles within the framework of the multisphere Mie scattering theory using the simplest dipolar-like approach shows that (1) high quality modes in an array of *10* or more particles can be attained at least for a refractive index $n_r > 2$, so optical materials like $TiO_2$ or GaAs can be used; (2) the most bound modes have nearly transverse polarization perpendicular to the circular plane; (3) in a particularly interesting case of $TiO_2$ particles (rutile phase, $n_r = 2.7$), the quality factor of the most bound mode increases almost by an order of magnitude with the addition of *10* extra particles, while for particles made of GaAs the quality factor increases by almost two orders of magnitude with the addition of ten extra particles. The consideration of higher multipole contributions has demonstrated that the error of the dipolar approach does not exceed one percent if the refractive index $n_r$ is greater than *2*. Minimum acceptable disordering not affecting the quality factor is studied.

## 1. INTRODUCTION

Scattering of light by objects of a size comparable to the wavelength of the light leads to a strong interference.[1] This effect is caused by the wave nature of light. Transmission of light through an inhomogeneous medium with the characteristic inhomogeneity size comparable to the wavelength is similar to electronic transport in condensed matter. This similarity is exploited in photonic crystals made of three dimensional ordered arrays of particles.[2,3] Particularly, one can create a photonic band-gap which would be impossible to accomplish using conventional optics. Photonic crystals having the forbidden optical band in all directions can be used to create localized defect states of light.[4] In a sufficiently large sample, these defect states formed by resonance defect modes within the forbidden band serve as resonant cavities with extremely narrow resonance characterized by the width decreasing exponentially with the size of the sample. Photonic crystals can be used in a variety of applications including channel-drop filters,[5] interconnects for optical circuits[6,7] and photonic crystal fibers.[8]

It is very difficult to make a perfect three dimensional photonic crystal with a period comparable to an optical wavelength because of problems with particle placement and structural disordering. Therefore, low-dimensional optical systems are attracting a significant attention.[9] The simplest low-dimensional realization is represented by one-dimensional structures, which we study in this work. It is interesting to note that even a single periodic chain of particles can possess high quality optical modes. This is due to the formation of a guiding mode bound to the chain as demonstrated by Ehrenspeck and Poehler in 1959 for the linear chain of coupled Yagi antennas.[10] Indeed, consider the periodic infinite chain of scattering particles (Fig. 1) with a size (diameter) *d*, a period *a* and a resonant frequency $\omega_0$ corresponding to a scattering (Mie) resonance for dielectric particles or a surface plasmon resonance for metal particles.[1] Each optical mode with the given frequency $\omega \approx \omega_0$ describing propagating waves can be characterized by its amplitude *A(z)* (electric or magnetic field) representing the solution of the Maxwell equations, where *z* is the coordinate along the chain (see Fig. 1). According to Bloch's theorem[11] in the system with period $a$, this solution can be chosen in the form satisfying the periodic condition

$$A(z+a) = A(z) \cdot e^{iqa}, \qquad (1)$$

where the parameter $q$ stands for the quasi-wavevector of the wave. Since the shift of the quasi-wavevector by the inverse chain period $q \rightarrow q+2\pi/a$ does not change Eq. (1), wavevectors differing by the integer number of inverse periods are equivalent. Therefore, one can restrict the domain of wavevectors to the elementary cell of the inverse lattice[11]

$$-\pi/a < q < \pi/a. \qquad (2)$$

---

[*] aburin@tulane.edu; phone 1 504 862-3574; fax 1 504 865-5596; tulane.edu

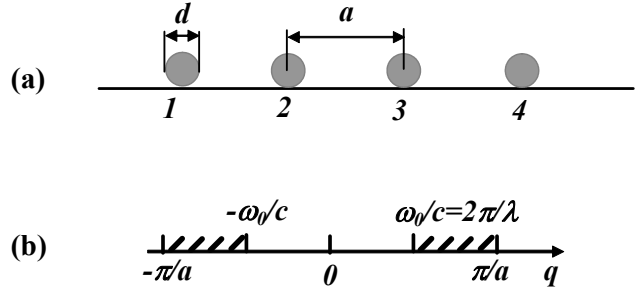

Fig. 1. Linear chain of scattering particles (a) and possible values of the quasi-wavevector of propagating polariton modes with marked domain of guiding modes (b). All notations are described within the text.

Generally, one can expect that the solution $A(z)$ occupies three-dimensional space because the volume including the chain has zero measure compared to the whole volume. At a significant distance from the chain the solution should exist in the form of a free photon with wavevector $k=\omega/c$, where $c$ is the speed of light. Since this solution ought to satisfy Eq. (1) anywhere in space, the $z$-projection of the free photon wavevector must be equal to the quasi-wavevector $q$ or differ from it by the integer number of inverse chain periods $2\pi/a$. For our choice of the domain of vectors $q$ we assume that the free photon wavevector $k$ is greater or equal to the quasi-wavevector $q$ of our mode. Therefore, if the wavevector $q$ exceeds the resonant wavevector $k$ of the free photon

$$q > k = \frac{\omega_0}{c}, \qquad (3)$$

the wave cannot escape far from the chain because of the breakdown of the momentum conservation law. If the resonant photon wavevector $q \approx \omega_0/c$ is less than the maximum quasi-wavevector $\pi/a$ (Eq. (2)), then there exist two spectral domains

$$\left(-\frac{\pi}{a}, -\frac{\omega_0}{c}\right), \left(\frac{\omega_0}{c}, \frac{\pi}{a}\right), \qquad (4)$$

where the optical modes are bound to the chain. These domains exist only if the resonant frequency is small enough ($\omega_0/c < \pi/a$) which is equivalent to the well known criterion that the period of the chain must be less than half of the resonant wavelength[10]

$$a < \frac{\pi c}{\omega_0} = \frac{\pi}{k} = \frac{\lambda_0}{2}. \qquad (5)$$

Thus, a one-dimensional infinite periodic chain of identical particles can possess optical modes which do not decay radiatively if the particles are located close enough to each other. Note that particles cannot be placed closer to each other than their characteristic size $d$. This raises the question as to whether Eq. (5) can be satisfied for particular optical material used to make scattering particles. Indeed, Eq. (5) can be easily satisfied for metal particles having a surface plasmon resonance.[1,12] In metal particles as small as *30* nm, this resonance belongs to the visible light spectrum corresponding to wavelengths as large as several hundred nanometers. Therefore, metal particles can be used to build waveguides free of radiative losses and such a waveguide was recently designed to transfer optical energy on a sub-wavelength scale.[13,14]

Nevertheless, all metals absorb optical energy because their conducting electrons have continuous spectrum and can consume arbitrary amount of energy. For gold and silver in particular, this absorption strongly reduces the energy transmission in nano-waveguides.[13,14] Therefore, it can be much more convenient to use dielectric materials having very weak light absorption. Instead of relying on surface plasmon resonance, one can use geometric scattering resonances to form polariton modes made of oscillations of material polarizations coupled to electromagnetic waves. In this paper we consider the optical system composed of a one dimensional circular array of dielectric spherical particles.

We can then attempt to create particle arrays possessing bound modes Eq. (5) using typical optical materials like $TiO_2$, ZnO or GaAs.[15] In contrast with metals, refractive indices of dielectric materials are nearly frequency independent; therefore, the resonant wavelength there is proportional to the particle size. At a large refractive index, the resonant wavelength also increases with the refractive index so we can express it as $\lambda = d f(n_r)$ where $f(n_r)$ is the increasing function of the refractive index $n_r$. At optical frequencies all dielectric materials have refractive indices of order unity (see Table 1) so $f(n_r) \sim 1$. Since the particle diameter represents the minimum possible period of the particle chain, Eq. (5) can be satisfied only if $f(n_r) > 2$.

Table 1. Quality factors of dielectric particles of different materials. The results are given only for one transverse mode $t1$, because the other transverse mode $t2$ behaves quite similarly in the limit of large number of particles $N$.

| Material | GaAs | | $TiO_2$ (rutile) | | ZnO | |
|---|---|---|---|---|---|---|
| $n_r$ | 3.5 | | 2.7 | | 1.9 | |
| Mode | $t1$ | $l$ | $t1$ | $l$ | $t1$ | $l$ |
| $Q(N=10)$ | 1700 | 1500 | 155 | 27 | 9 | 4 |
| $Q(N=50)$ | $1.8 \cdot 10^{10}$ | $4.77 \cdot 10^7$ | $8.42 \cdot 10^5$ | 7000 | 34.2 | 19 |
| $d\ln Q/dN$ (num., dip.) | 0.4025 | 0.3129 | 0.21 | 0.1304 | 0.0153 | 0 |
| $d\ln Q/dN$ (analyt.) | 0.39 | 0.2965 | 0.202 | 0.12 | 0.0342 | 0 |
| $k$ ($a=d=2$) | 0.84337 | 0.945 | 1.06824 | 1.2035 | 1.4042 | 1.598 |
| $\sigma(N=20)$ | 0.005 | n/a | 0.05 | n/a | n/a | n/a |
| $\sigma(N=50)$ | <0.00001 | n/a | 0.0015 | n/a | n/a | n/a |

In this paper we show that long-living modes possessing a high quality factor can indeed be created using a few particles made of standard optical materials with refractive indices greater than *2.5*. Following the preliminary publication,[16] we consider circular arrays of particles as one-dimensional objects of highest symmetry which are expected to possess a very high quality factor that increases exponentially with the number of scattering particles.[16,17] In this paper we study modes having the lowest frequency (those related to the first Mie resonance), which are most easily bound because they possess the largest wavelength (see Eq. (5)).

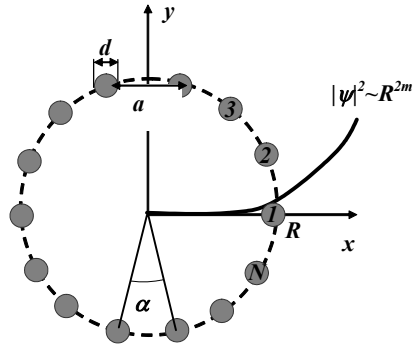

Fig. 2. Circular array of scattering particles. The radiative losses of the whispering gallery mode belonging to this array are defined by its exponentially small overlap with the photon possessing the same angular momentum projection *m* to the z-axis (see text for details).

We investigate the whispering gallery modes within a one-dimensional ring of particles (Fig. 2) using the multi-sphere Mie scattering formalism earlier developed by Y. L. Xu.[18,19] This formalism is equivalent to multipole expansion of electric and magnetic fields of each sphere with further investigation of interactions between them. This formalism has a distinct advantage compared to the standard finite difference time domain (FDTD) approach[20] in that it reduces the hard problem of the solution of Maxwell equations in the whole space to the analysis of a discrete set of parameters, i. e. the multipole moments of the scattering particles. The solution of the problem is valid everywhere in space and therefore the result is independent of boundary conditions which can critically affect the solution obtained using the FDTD method. As we proved here, the analysis of the quality factor (not necessarily field amplitudes) of low energy optical modes for reasonably large refractive index ($n_r>2$) can be performed using only dipolar approach, which makes numerical calculations fast therefore permitting us to analyze the large arrays of hundreds or even thousands of particles.

This paper is organized as follows. In Section 2, we present the qualitative analysis of the quality factor of modes within the circular array of dielectric particles leading to the exponential increase of the quality factor with the number of particles earlier predicted for circular arrays of antennas.[21] In Section 3, we describe the multi-sphere Mie scattering formalism and its application to the investigation of eigensolutions of Maxwell equation. In Section 4, the combined analytical and numerical solution for ordered circular arrays of identical particles is obtained in the dipolar approach. We also discuss the numerical results for quality factor dependence on the number of particles and corrections to the dipolar approach induced by the higher multipole moments. In addition, we consider the disordering effect on the quality factor. The minimum acceptable disordering which does not affect the quality factor for the given refractive index and particle number is estimated. In Section 5 our conclusions are given.

## 2. QUALITATIVE ANALYSIS OF BOUND MODES

Here we suggest a simple derivation for the quality factor of a circular array of dielectric spherical particles shown in Fig. 2, which reproduces earlier results for Yagi arrays.[17,21] Assume that this array consists of $N$ particles separated from each other by the distance $a$ and placed periodically into the circle of radius

$$R \approx aN/(2\pi). \tag{6}$$

The particle array has a rotational symmetry. The rotation of the whole system by the integer number of the angle

$$\alpha = 2\pi/N \tag{7}$$

with respect to the center of the circle moves the system to its initial orientation (see Fig. 2). This symmetry permits us to use the analog of the Bloch theorem Eq. (1) for the angular dependent optical mode amplitude $A(\varphi)$ in the form

$$A(\varphi+\alpha) = A(\varphi) \cdot e^{im\alpha}, \tag{8}$$

where $\varphi$ is the azimuth angle used in spherical coordinates and $m$ is the integer number representing the projection of the mode angular quasi-momentum onto the $z$-axis. The requirement for $m$ to be an integer is the consequence of the coordinate invariance after rotation by angle $2\pi$ within the $x$-$y$ plane. Since the change $m \to m+N$ does not affect Eq. (8), one can restrict possible values of the angular momentum projection $m$ by the domain $(-N/2, N/2)$ if $N$ is an even number or $(-(N-1)/2, (N+1)/2)$ if $N$ is an odd number. For simplicity, we restrict our consideration to only even numbers $N$ so that $N/2$ is an integer number.

Consider the radiative decay of the whispering gallery mode possessing a certain value of the angular momentum projection $m$. Assuming that this mode is nearly localized within the circular array of particles, one can define its decay rate using its overlap $\psi_m(R)$ with the wavefunction $\psi_m(r)$ of the free photon possessing the same angular momentum projection. The angular momentum projection $m$ corresponds to a total angular momentum $l \geq m$. Since at a small radius $R$ the photon wavefunction behaves as $R^l$ (Fig. 2), one can expect the exponential reduction of the polariton mode decay rate with increasing its angular momentum projection $m$. With a more accurate approach, one can use cylindrical coordinates where the photon wavefunction radius dependence is represented by the Bessel function[22] $\psi_m(r) \propto J_m(kr)$, where $k=\omega_0/c$ is the wavevector of the resonant photon having the polariton frequency $\omega_0$. The whispering gallery mode decay rate can be estimated as the squared photon wavefunction taken at the position of the array $\gamma_m \propto |J_m(kR)|^2$. The mode with the maximum angular momentum projection $m=N/2$ possesses a minimum decay rate because at large index $m$, the Bessel function always decreases with the increase of its index.[23] The dependence of the minimum decay rate $\gamma_{N/2}$ on the number of particles can be found using the expression Eq. (6) in the form

$$\gamma(N) \propto J_{N/2}^2(Nka/(2\pi)).  \qquad (9)$$

If the interparticle distance $a$ exceeds half the resonant wavelength ($ka/\pi > 1$), then the decay rate decreases exponentially with the number of particles $N$ as[23]

$$\gamma(N) \propto e^{-\kappa N}, \kappa = -\frac{d\ln\gamma}{dN} = \left[\cosh^{-1}\left(\frac{1}{x}\right) - \sqrt{1-x^2}\right], x = \frac{ka}{\pi}. \qquad (10)$$

It is clear that the quality factor $Q=\omega/(2\gamma)$ increases with increasing $N$ as $e^{\kappa N}$. This result reproduces the results of Refs. [17, 21] obtained for circular arrays of cylindrical antennas, and it generalizes these previous results to optical systems. Later in this manuscript we will see that Eqs. (9) and (10) are in agreement with the numerical calculations of whispering gallery modes in circular arrays of spherical particles (see Table 1).

Note that it can be possible to make the non-radiating current formally corresponding to the mode having an infinitely high quality factor.[24,25] Due to symmetry, this mode can have no multipole moments, for instance no light can be emitted for the spherically symmetrical charge distribution. However, to our knowledge, the real example of such a mode given in Ref. [25] uses an infinitely long cylinder and it is not clear how to create a system with a non-radiating polariton mode within a finite volume in three dimensions.

## 3. MULTI-SPHERE MIE-SCATTERING FORMALISM

In this work we use the multisphere Mie scattering formalism earlier developed by Y. L. Xu[18,19] (see also Ref. [26]) to study the quality factors of whispering gallery modes in arrays of spherical particles placed along the ring. This multisphere Mie scattering formalism has been developed to study the scattering of light in aggregates of spherical dielectric particles. The formalism uses the spherical vector function expansion of solutions to Maxwell's equations in the frequency domain taken at frequency $z$. The scattering wave partial amplitudes $a_{mn}^l$, $b_{mn}^l$ ($l=1, 2, ...N$) describing two transverse components with angular momentum $n$ and its projection to the z-direction $m$ all taken with respect to the center of the $l^{th}$ sphere are expressed through partial amplitudes $p_{mn}^l$ and $q_{mn}^l$ and by making use of matrices $A$ and $B$ defined by the vector translation coefficients (see Refs. [18, 19] for details)

$$\frac{a_{mn}^l}{\overline{a}_n^l} + \sum_{j \neq l}^{(1,N)} \sum_{\nu=1}^{+\infty} \sum_{\mu=-\nu}^{\nu} \left(A_{mn\mu\nu}^{jl} a_{\mu\nu}^j + B_{mn\mu\nu}^{jl} b_{\mu\nu}^j\right) = p_{mn}^l,$$

$$\frac{b_{mn}^l}{\overline{b}_n^l} + \sum_{j \neq l}^{(1,N)} \sum_{\nu=1}^{+\infty} \sum_{\mu=-\nu}^{\nu} \left(B_{mn\mu\nu}^{jl} a_{\mu\nu}^j + A_{mn\mu\nu}^{jl} b_{\mu\nu}^j\right) = q_{mn}^l, \qquad (11)$$

where $\overline{a}_n^l$, $\overline{b}_n^l$ are Mie scattering coefficients for the $l^{th}$ sphere.[1,18,19] Matrices $A$ and $B$ describe the multipole interactions. Case $n=1$ corresponds to a dipole moment, case $n=2$ describes a quadrupole moment, etc.

Since we are interested in eigenmodes of a finite system, the left hand side of Eq. (11) describing the source term should be set to zero. This is equivalent to the analysis of quasistates[15] defined by solutions of homogeneous Maxwell equations for the finite system of scattering particles with the boundary condition to have only outgoing waves at an infinity. Such solutions can exist at a discrete set of frequencies having an imaginary part because of the "dissipative" boundary conditions.

In addition, considering that we cannot treat an infinite number of indices $m$ and $n$ in the expansion of the wave, one should choose some upper constraint $n_{max}$ on the maximum angular momentum. The choice of $n_{max}$ depends on the frequency $\omega$ of the mode of interest. Since we are interested in the modes most strongly bound to the array of particles, the frequency of the mode should be selected to be as small as possible to better satisfy the guiding criterion Eq. (5). The lowest resonance corresponds to dipolar scattering so the most relevant scattering and interaction corresponds to indices $n$, $\nu=1$ in Eq. (11). Therefore, the most straightforward approach is to set $n_{max}=1$, which is equivalent to the dipolar approach.[27] This was done in Ref. [16] and here we report these results in greater details. In addition, we verify the applicability of the dipolar approach by increasing $n_{max}$.

In the dipolar approximation we also ignore the off-diagonal interaction **B** between the $a_{m1}^l$ and $b_{m1}^l$ components. It follows that we have created two separate sets of equations for amplitudes **a** and **b** involving only the diagonal interaction matrix **A**

$$\frac{a_{m1}^l}{\bar{a}_1^l} + \sum_{j \neq l}^{(1,N)} \sum_{\mu=-1}^{1} A_{m1\mu1}^{jl} a_{\mu1}^j = 0, \tag{12.a}$$

$$\frac{b_{m1}^l}{\bar{b}_1^l} + \sum_{j \neq l}^{(1,N)} \sum_{\mu=-1}^{1} A_{m1\mu1}^{jl} b_{\mu1}^j = 0. \tag{12.b}$$

Recall that Mie scattering coefficients are defined by Ricatti-Bessel functions $\psi_1(x) = \sin(x)/x - \cos(x)$ and $\varsigma_1(x) = e^{ix}(-1 - i/x)$ as

$$\begin{aligned}
\bar{a}_1 &= \frac{\psi_1(kd/2)\psi_1'(n_r kd/2) - n_r \psi_1(n_r kd/2)\psi_1'(kd/2)}{\varsigma_1(kd/2)\psi_1'(n_r kd/2) - n_r \psi_1(n_r kd/2)\varsigma_1'(kd/2)}, \\
\bar{b}_1 &= \frac{n_r \psi_1(kd/2)\psi_1'(n_r kd/2) - \psi_1(n_r kd/2)\psi_1'(kd/2)}{n_r \varsigma_1(kd/2)\psi_1'(n_r kd/2) - \psi_1(n_r kd/2)\varsigma_1'(kd/2)}.
\end{aligned} \tag{12.c}$$

The nontrivial solutions of Eq. (12) exist for a discrete sequence of eigenfrequencies $z_a = \omega_a + i\gamma_a$. We are interested in finite systems where the emission of photons is inevitable so each eigenmode possesses the finite decay rate $\gamma_a$. Our consideration starts with the analysis of Eq. (12). This equation defines two spectral branches corresponding to Mie resonances associated with the zeros of inverse Mie scattering amplitudes $1/\bar{a}_1^l$ and $1/\bar{b}_1^l$. Since the second amplitude has a resonance at a lower frequency (cf. Ref. [1]), it is of greatest interest to us and we will study the modes associated with the solution of Eq. (12.b).

The study will proceed as follows. First we will investigate Eq. (12.b) for lowest energy whispering gallery modes in a circular array made of identical, equally separated dielectric spheres composed of various optical materials of interest represented in Table 1. Then the accuracy of the dipolar approach will be checked by the additional consideration of the off-diagonal interaction and higher multipoles. Then the disordering effect on the quality factor of most bound polariton modes will be addressed. Below we describe the method of our study which incorporates the modified Newton-Raphson method in determining solutions to the transcendental equations. This method is partially inherited from Refs. [15, 28], where lasing modes were investigated for interacting dipolar oscillators.

For an array of *N* ordered or disordered spheres Eq. (12.b) can be rewritten as

$$\hat{M}(z)\mathbf{b} = 0, \tag{13}$$

where **M** is the matrix $3N \times 3N$ extracted from Eq. (12.b). The diagonal elements of this matrix are inverse Mie scattering coefficients and its off-diagonal elements are defined by the matrix **A** of dipolar interactions between polarizations of different spheres. Vector **b** represents $3N$ expansion coefficients $b_{m1}^l$. The nontrivial solution of Eq. (13) exists, when the matrix **M** has a zero eigenvalue. We need to find the value of *z* which makes one of the eigenvalues of matrix *M(z)* zero and possesses the minimum imaginary part. To use the Newton-Raphson algorithm we need to choose the initial value of frequency. This value can be chosen as the Mie resonance frequency for the typical sphere. Then we define the function *f(z)* as some eigenvalue of the matrix *M(z)* (cf. [15, 28]) that minimizes the absolute value of the imaginary part of *z* in the next step of iteration defined in the standard way

$$z_{n+1} = z_n - \frac{f(z_n)}{df(z_n)/dz}. \tag{14}$$

The stable point of this algorithm is realized when *f(z)=0*. Then Eq. (13) and consequently Eq. (12.b) have the nontrivial solution because the matrix **M** has one zero eigenvalue. Our method of iteration has been chosen to find the minimum eigenvalue, although we cannot prove rigorously that it always converges to the right eigenfrequency. Eq. (12.b) can be greatly simplified using the Fourier transform methods Eqs. (1), (8) for identical equally spaced particles, arranged as an infinite chain (Fig. 1) or a finite circle (Fig. 2). Then $3N \times 3N$ matrix $M$ in Eq. (13) is replaced by a $3 \times 3$ matrix of Fourier transforms of off-diagonal components from the original matrix $M$ (see Sec. 4 for details). Then our solution can easily be verified comparing solutions obtained by different methods. It happens that if the initial frequency $z_0$ in Eq. (14) is chosen to be real and have the real part smaller that the lowest Mie resonance frequency, then for a reasonable number of particles ($N \leq 70$) and a refractive index $n_r > 2.5$, the iteration procedure Eq. (14) rapidly converges in less than *10* steps to the right value. For a larger number of particles, a smaller refractive index, and appreciable disordering in particle sizes and placements, the procedure also results in the right value if there is convergence. Otherwise the iteration series can be trapped by an attractor,[29] which is the unlimited self-repetition of sequence of few (two or three) points instead of the convergence to the single point. In the case of an attractor we restart the program with a different initial value $z_0$ until we have convergence. Our choice of $z_0$ with a large imaginary part or with a real part exceeding the Mie resonance frequency has often led to a different solution possessing a large decay rate, which is not relevant for our study.

The problem dealing with the extended Eq. (12.b) and added multipole interactions to the matrix $M$ in Eq. (13) can be generally expressed in the form

$$\begin{bmatrix} \hat{M}(z) & \hat{M}_{12}(z) \\ \hat{M}_{21}(z) & \hat{M}_{22}(z) \end{bmatrix} \begin{pmatrix} \mathbf{b} \\ \mathbf{b}_2 \end{pmatrix} = 0, \tag{15}$$

where block-matrices $M_{12}$, $M_{21}$ and $M_{22}$ stand for higher multipole interactions and vector $b_2$ stands for amplitudes of higher multipole components. Our study shows that it is possible to account for the effect of higher multipoles by excluding vector $b_2$ from consideration by means of the replacement $b_2 = -M_{22}^{-1} M_{21} b$ to solve Eq. (13) with matrix $M$ modified as

$$\hat{M}' = \hat{M} - \hat{M}_{12} \hat{M}_{22}^{-1} \hat{M}_{21} \tag{16}$$

and applying the Newton-Raphson method Eq. (14) as described above. The direct analysis of the large block matrix in Eq. (15) leads to solutions corresponding to higher frequency modes than the one we are seeking.

## 4. INVESTIGATION OF CIRCULAR ARRAYS

We begin our study with the analysis of the quality factor of ordered circular arrays of identical equally spaced particles within the framework of the dipolar approach Eq. (12.b). Then the effect of multipole interaction is studied and the effect of disordering is discussed.

We begin our analysis with the consideration of whispering gallery modes within the circular array shown in Fig. 2 making use of the dipolar approach Eq. (12.b). The Mie resonances that form these modes are selected based on having the lowest frequency compared to other resonances. A more accurate approach including consideration of different interacting resonances can lead to the reduction of the resonant frequencies of the modes of interest similarly to the quantum mechanical energy level repulsion effect, where the off-diagonal perturbation always reduces the ground state energy.[16] For a large number of particles $N$, this effect leads to an increase of the mode quality factor because the mode decay rate Eqs. (9), (10) decreases with decreasing the resonant photon frequency $\omega$ and its resonant wavevector $k=\omega/c$. Therefore the dipolar approach overestimates decay rates and underestimates quality factors for modes of interest.

One can formally distinguish whispering gallery modes by their polarization. For a chain of particles (Fig. 1), one can distinguish two transverse modes: two modes with polarizations directed perpendicular to the chain and one longitudinal mode with polarization directed parallel to the chain. For large $N$ a circular array behaves similarly to the chain of particles so one can approximately expect to have three distinguishable modes including the first transverse mode (*t1*) with the polarization perpendicular to the circle *x-y* plane, i.e. along the z-axis; the second transverse mode (*t2*) having the polarization along the circle radius; and the longitudinal mode (*l*) with the polarization approximately tangential to the circle. The mode *t1* is described by the angular momentum projection $m=0$, i. e. amplitudes $b_{01}$, while modes *t2* and *l* are described by the superposition of amplitudes $b_{11}$ and $b_{-11}$. The mode *t1* is not coupled with modes *t2* and *l* because the electric field of the dipole directed perpendicularly to the *x-y* plane and taken at the point belonging to the *x-y* plane is also perpendicular to the x-y plane. Therefore the dipoles polarized along the z-direction interact only with the dipoles having the same polarization. Then Eq. (12.b) for the mode *t1* can be simplified to

$$\frac{b_{01}^l}{\bar{b}_1^l} + \sum_{j \neq l}^{(1,N)} A_{0101}^{jl} b_{01}^j = 0 \tag{17}$$

while equations for modes *t2* and *l* take the form

$$\frac{b_{11}^l}{\bar{b}_1^l} + \sum_{j \neq l}^{(1,N)} \left( A_{1111}^{jl} b_{11}^j + A_{11-11}^{jl} b_{-11}^j \right) = 0, \quad \frac{b_{-11}^l}{\bar{b}_1^l} + \sum_{j \neq l}^{(1,N)} \left( A_{-11-11}^{jl} b_{-11}^j + A_{-1111}^{jl} b_{11}^j \right) = 0. \tag{18}$$

The dielectric spheres are enumerated in the counterclockwise direction as shown in Fig. 2. Eqs. (17), (18) can be simplified using the Fourier transform method. This can be done using the dependence of the matrix elements $A$ on distance $r$ between centers of spheres expressed in terms of the dimensionless product $x=kr$ (remember that $k=\omega/c$ is the resonant photon wavevector) and the angle $\varphi$ between the vector connecting two spheres and the *x*-axis, which can be represented in the form of dipolar fields[16,18,19]

$$A_{0101}(x,\varphi) = 1.5i \frac{e^{ix}}{x}\left(-1 - \frac{i}{x} + \frac{1}{x^2}\right), \quad A_{1111}(x,\varphi) = A_{-11-11}(x,\varphi) = 0.75i \frac{e^{ix}}{x}\left(-1 - \frac{i}{x} - \frac{1}{x^2}\right),$$

$$A_{-1111}(x,\varphi) = i \frac{3e^{ix+2i\varphi}}{8x}\left(-1 + \frac{3i}{x} + \frac{3}{x^2}\right), \quad A_{11-11}(x,\varphi) = i \frac{3e^{ix-2i\varphi}}{2x}\left(-1 + \frac{3i}{x} + \frac{3}{x^2}\right). \tag{19}$$

The difference of a factor of four in the definition of matrix elements $A_{-1111}$ and $A_{11-11}$ is caused by the definition of Legendre polynomials.[16,18] Using Eq. (19) one can express the solutions of Eqs. (17), (18) in the form[16] (cf. Eq. (8))

$$b_{m1}^l = e^{i2\pi(M+m)l/N}, \tag{20}$$

where the integer index $M=0, 1, \dots N-1$ enumerates the possible projection of the mode quasi-angular momentum onto the *z*-axis. Substituting Eq. (20) into Eq. (18) and requiring the existence of a non-trivial solution we obtain equations defining the spectra of all modes in terms of Fourier transforms of various interactions

$$F_k(M,q,N) = \sum_{l=1}^{N-1} \frac{e^{2iqR\sin(l\pi/N)+2\pi iMl/N}}{(2qR\sin(l\pi/N))^k},$$

where $R$ is the radius of the circle.

For a mode *t1* we get

$$\frac{1}{\bar{b}_1} + \varepsilon_0(M,qa,N) = 0,$$

$$\varepsilon_0(M,qa,N) = -\frac{3}{2}iF_1(M,qa,N) + \frac{3}{2}F_2(M,qa,N) + \frac{3}{2}iF_3(M,qa,N) = 0. \tag{21}$$

The spectrum of a *t2* mode is defined by the following equation[16]

$$\frac{1}{\bar{b}_1} + \frac{1}{4}\left(\varepsilon_+ + \sqrt{\varepsilon_-^2 + V^2}\right), \varepsilon_\pm = \frac{\sigma(k,M+1) \pm \sigma(k,M+1)}{2},$$

$$\sigma(k,M) = \frac{3}{4}\left(\Sigma_1(M,qa,N) - i\Sigma_2(M,qa,N) + \Sigma_3(M,qa,N)\right) \tag{22}$$

and the spectrum of a longitudinal mode is defined as

$$\frac{1}{\bar{b}_1} + \frac{1}{4}\left(\varepsilon_+ - \sqrt{\varepsilon_-^2 + V^2}\right). \tag{23}$$

Eqs. (21), (22), (23) can be resolved using the Newton-Raphson method or a generalization of the iteration method developed in Ref. [28]. All methods converge to the same frequency having the lowest imaginary part at a particular value of the angular momentum projection $M$. We are interested in the most strongly bound modes possessing the smallest decay rate (i. e. the highest quality factor). Our study shows that in the regime of interest where the quality factor exceeds $10$ this mode is formed at $M=N/2$, which corresponds to the maximum value of the quasi-momentum $q=\pi/a$ (Eq. (4)). Indeed, the angular momentum projection is given by $M=qR=(\pi/a)\cdot Na/(2\pi)=N/2$. Also it is clear that this regime of most strongly bound modes is realized when the particles are closest to each other. This takes place when the distance between the centers of particles is equal to their diameter. Therefore in future study we set

$$M = N/2, \quad d = a. \tag{24}$$

In this particular case we will also consider the contribution of non-dipolar interactions. We consider corrections associated with the interaction of the first Mie resonance with second, fourth and fifth Mie resonances, The Fourier transform method can be used to simplify equations with $a_{mn}^l$ and $b_{mn}^l$ ($n=1, 2, 3$) components taken in the form Eq. (20). We restrict our consideration to the transverse mode $t1$ possessing the highest quality factor. In the special case of Eq. (24) (remember that we are interested only in the case of even number of particles $N$) the corrected spectrum of $t1$ mode Eq. (21) with $M=N/2$ can be expressed as (the limit $N>>1$ is also assumed)

$$\frac{1}{\bar{b}_1} + \varepsilon_0(M, qa, N) - \frac{4b_{0111}(N/2-1/2)^2}{\frac{1}{\bar{a}_1} + a_{11}(N/2-1) - 2a_{-11}(N/2)} - \frac{b_{0102}(N/2)b_{0201}(N/2)}{\frac{1}{\bar{a}_2} + a_{2121}(N/2)}$$
$$- \frac{a_{0112}(N/2-1/2)a_{1201}(N/2+1/2)}{\frac{1}{\bar{b}_2} + a_{1212}(N/2-2)} - \frac{a_{01-12}(N/2+1/2)a_{-1201}(N/2-1/2)}{\frac{1}{\bar{b}_2} + a_{-12-12}(N/2-1)} -$$
$$- \frac{b_{0122}(N/2-1)b_{2201}(N/2+1)}{\frac{1}{\bar{a}_2} + a_{2222}(N/2-1)} - \frac{b_{01-22}(N/2+1)b_{-2201}(N/2-1)}{\frac{1}{\bar{a}_2} + a_{-22-22}(N/2-1)} - \frac{a_{0103}(N/2)a_{0301}(N/2)}{\frac{1}{\bar{b}_2} + a_{0303}(N/2)} -$$
$$- \frac{a_{0123}(N/2-1)a_{2301}(N/2+1)}{\frac{1}{\bar{b}_3} + a_{2323}(N/2-1)} - \frac{a_{01-23}(N/2+1)a_{-2301}(N/2-1)}{\frac{1}{\bar{a}_2} + a_{-23-23}(N/2-1)} = 0, \tag{25}$$

$$b(a)_{mn\mu\nu}(M) = \sum_{j=2}^{N} B(A)_{mn\mu\nu}^{0j} e^{i2\pi Mj/N - i(m-\mu)(\pi/2 - \pi j/N)}.$$

There is no coupling between several modes because of the symmetry of the system[18-19] (vanishing in-plane interactions are $A_{0102}= A_{0122}= A_{01-22}=A_{0113}=A_{01-13}= A_{0133}= A_{01-33}=B_{0101}=B_{0112}=B_{01-12}=0$) which are not included into Eq. (25).

The effect of interaction of the lowest Mie resonance with higher Mie resonances will be studied within the framework of four approximations. The first approximation is the dipolar approach where only two first terms are left in Eq. (25). In the second approximation ($n=1$) we add the third term (off-diagonal coupling with the amplitude $a_{01}$). In the third approximation ($n=1, 2$) first eight terms in Eq. (25) are considered involving dipole-quadrupole interactions. In the fourth approximation ($n=1, 2, 3$) interactions with all five Mie resonances are considered. These approximations can be treated as the perturbation series. The strong convergence of this series is illustrated for several materials in Fig. 4.

In numerical calculations, we have chosen interparticle distance $a=2$. Since mode frequencies and decay rates are inversely proportional to the particle size one can easily recalculate them for any size $a$ of interest. The results are presented using the size independent quality factor, which is defined as half of the ratio of real and imaginary parts of a mode eigenfrequency

$$Q = \frac{\text{Re}(z)}{2|\text{Im}(z)|} = \frac{\omega}{2\gamma}. \tag{26}$$

Our calculations of the quality factor for different refractive indices and different modes are shown in Figs. 3, 5 and 6. There is no difference between modes *t1* and *t2* for a large number of particles where the circular array behaves like an infinite chain as these two modes degenerate. Therefore, data are shown for the *t1* mode only, where the quality factor is slightly higher. For all materials quality factors of transverse modes exceed those for longitudinal modes.

It turns out that the modes with the highest quality factor are found using GaAs. This is not surprising because GaAs has the highest refractive index. Accordingly it has a Mie resonance at the lowest frequency so guiding criterion Eq. (5) can be most easily realized there. One should note that for only *N=10* particles the quality factor of the *t1* mode already exceeds *1000*, while if the number of particles exceeds *70* the quality factor reaches the astronomical value $\sim 10^{14}$. It is clear that one can form a bound whispering gallery mode using GaAs because the quality factor increases greatly with the number of particles. This behavior can be fit by the exponential dependence as shown in Fig. 3. For instance the *t1*-mode is characterized by the quality factor approximate behavior (see Fig. 2)

$$Q(N) = 32.53 \cdot \exp(0.4025 N). \tag{27}$$

The important characteristic of this dependence is the exponential growth rate $d\ln(Q(N))/dN \approx 0.402$ given by the fifth row of Table 1. It is interesting to compare the calculated exponential behaviors with the prediction Eq. (10). For modes of interest with the angular momentum projection *M=N/2*, we found resonant photon wavevectors to be $k_t = 0.84337$, and $k_l = 0.945$ (see Table 1, remember that we set *a=d=2*). Both for transverse and longitudinal modes these wavevectors are smaller than the maximum polariton quasi-wavevector $\pi/a = \pi/2$ as necessary for bound modes. Using the photon wavevectors in Eq. (10), one can estimate exponential growth rates as shown in the sixth row of Table 1. There is always good agreement between the result of calculations and the theory prediction for GaAs.

The reported results were obtained within the framework of the dipolar approximation. To estimate the accuracy of this approximation we take into account higher multipole interactions. The most significant effect is caused by the interaction of two lowest Mie resonances due to the off-diagonal interaction Eq. (25). Using Eq. (25) we calculated corrections to the frequency, decay rate and quality factor for the *t1*-mode in *GaAs* ($n_r = 3.5$). These corrections indicated by obscured star symbols in Fig. 3 because of their strong coincidence with the dipolar approximation. The results of our analysis of Eq. (25) are presented in Fig. 4.

As was discussed above and in Ref. [16], the corrections always reduce the resonant frequency because of the "level repulsion effect" between high and low frequency Mie resonances. The corrections to the frequency are less than 1%. The off-diagonal interaction with the adjacent Mie resonance (*n=1*) affects the resonant frequencies at small number of particles *N* while its effect disappears with increasing *N*. The interaction between two resonances vanishes in the limit of large *N*, but we cannot interpret this fact. Two non-vanishing corrections are associated with the second (*n=1, 2*) and third (*n=1, 2, 3*) resonances. The second correction is the largest one (as the first non-vanishing effect), while the third correction is much smaller than the second one. This indicates the fast convergence of the whole series of corrections. A quality factor shows similar dependence on the order of perturbation theory under consideration so we believe that our calculation is accurate enough.

The corrected quality factor deviates more strongly from the results of the dipolar approximation. This is understandable because the difference in mode frequencies results in the difference in exponential growing rate Eq. (10). Therefore, at large number of particles, the quality factor for the mode with a lower frequency is always much greater than that for the mode of a higher frequency. It is clear from Fig. 4 that the corrected quality factor increases with respect to the one calculated within the dipolar approach. However this increase is slow so all approximations remain comparable to each other. It is interesting that the corrections also affect the preexponential factor in the quality factor dependence on *N* (cf. Eq. (27)). The new preexponential factor is smaller by the factor of *1.5*. The nature of this reduction is not clear for us. Our conclusion about the applicability of the dipolar approximation is valid for frequencies, decay rates and quality factors, but it is not necessarily valid for electric and magnetic fields, which might show a singular behavior especially at the points where spheres touch each other.[30]

The quality factor was also calculated within the framework of the dipolar approach for two materials with lower refractive indices including TiO$_2$ ($n_r=2.7$, rutile phase) and ZnO ($n_r=1.9$). The results for TiO$_2$ are shown in Fig. 5. One can see that the behavior of the quality factor there is quite similar to GaAs with the only difference that it is smaller there because of the smaller refractive index. The quantitative characteristics of the particle array made of TiO$_2$ are given in Fig. 4 and included into table 1 similarly to those for GaAs. Still only ten particles are needed to attain the quality factor exceeding one hundred. The numerical results are in the reasonable agreement with the analytical estimate based on the mode overlap integral Eq. (10).

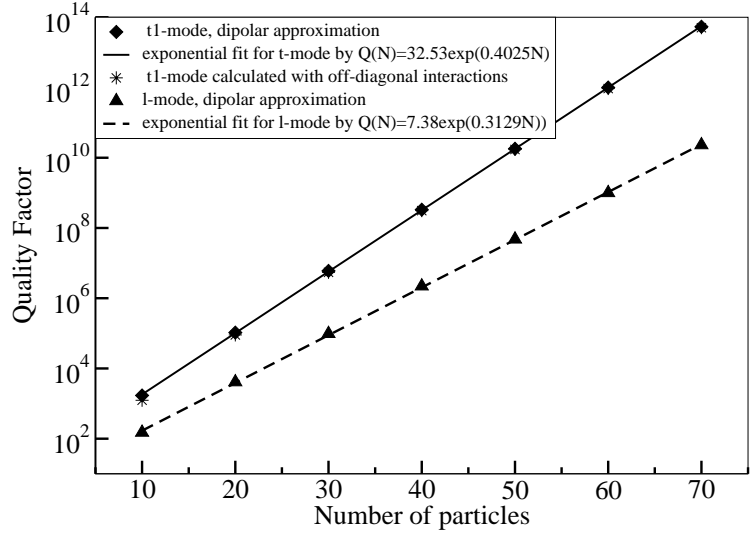

Fig. 3. Quality factor dependence on the number of particles for GaAs ordered circular arrays.

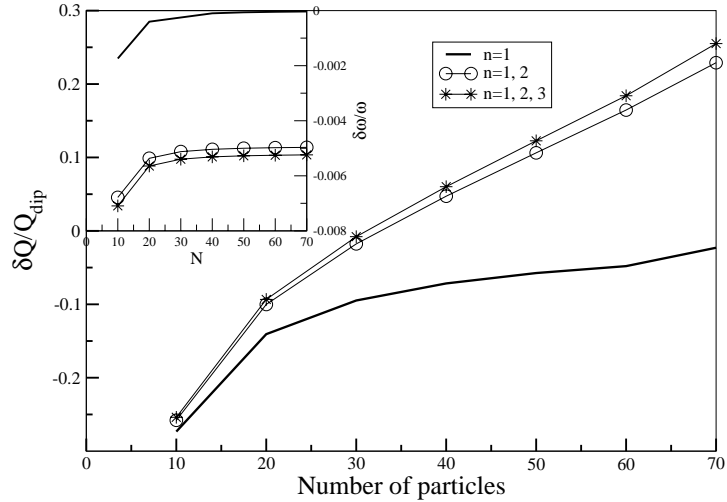

Fig. 4. Relative corrections to whispering gallery mode resonance frequency (inset) and quality factor due to multipole interactions for GaAs ($n_r=3.5$).

The behavior of the quality factor in ZnO differs qualitatively from two previous cases (see Fig. 6). This is because the refractive index of ZnO ($n_r=1.9$) is the smallest amongst the materials under consideration. Therefore, at least within the framework of the dipolar approximation, the only transverse whispering gallery mode is bound, while the longitudinal mode is unbound because its wavevector $k=1.598$ does not satisfy the guidance criterion Eq. (5). Accordingly at large number of particles $N$, the quality factor becomes independent of $N$ (Fig. 6). Since the difference of $k$ from the threshold is smaller than *2%* it is not clear whether all longitudinal modes are really unbound because the corrections to the frequency can be imported. The analysis of these corrections is outside the scope of this work. The transverse mode is bound, but it is very close to the threshold so the estimate for the logarithmic growing rate Eq. (10) is probably not applicable for sizes *N* used in our study (see Table 1).

The corrections to the *t1*-mode frequency and quality factor of $TiO_2$ and ZnO are quite similar to GaAs (Fig. 4). The absolute value of the correction to frequency is larger in ZnO and can reach *2-3%* of its absolute value. The convergence of the series of perturbations (*n=1; n=1,2; n=1, 2, 3*) is very good for all materials.

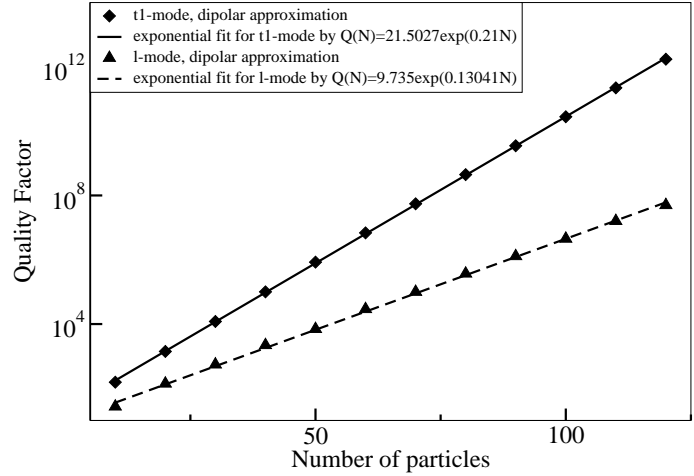

Fig. 5. Quality factor dependence on the number of particles for $TiO_2$ ordered circular arrays.

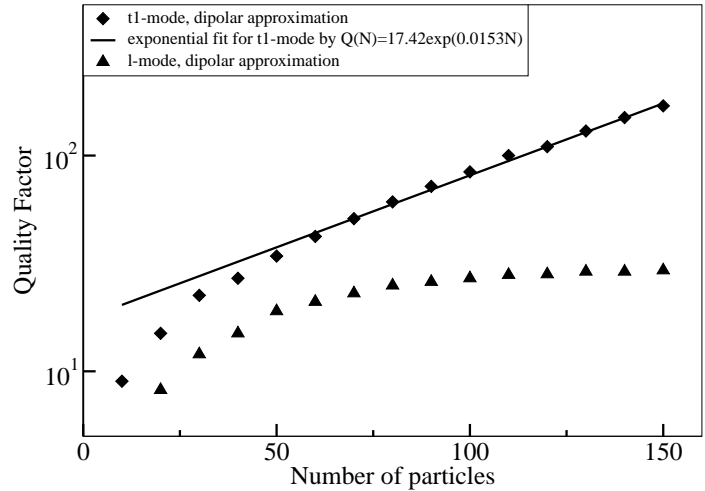

Fig. 6. Quality factor dependence on the number of particles for ZnO ordered circular arrays.

Finally we proceed to the discussion of disordering. In this paper we report the simple estimate of the maximum acceptable disordering necessary to reduce quality factor by the factor of 10. We introduced disorder in particle sizes generating the random correction to each particle diameter distributed normally with the certain dispersion $\sigma$. For the sake of simplicity, particles were placed along the circle in touch with each other. A characteristic minimum value of $\sigma$ leading to the reduction of the quality factor by the factor of $\xi$, such as ($5<\xi<20$), was found by probing random values of $\sigma$ until the first success. This crude method must be quite sufficient for qualitative characterization of disordering effect. The dependence of the parameter $\sigma$ on the number of particles is shown in Fig. 7 for transverse modes using different materials.

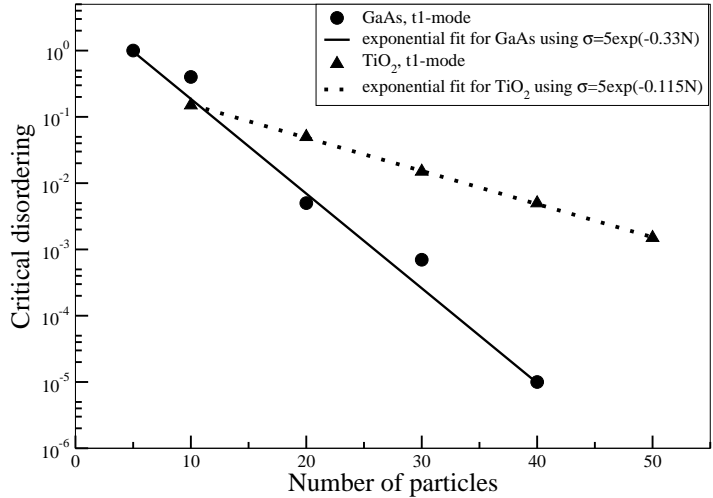

Fig. 7. Acceptable disordering for GaAs and TiO$_2$, calculated assuming that it reduces the quality factor by around factor of ten compared to the regular ring of particles.

According to our crude modeling presented in Fig. 7, the "critical" dispersion $\sigma$ shows an exponential dependence on the number of particles $N$. This is not surprising because the quality factor increases exponentially with $N$ and the sensitivity of the system to disordering should increase with increasing the quality factor. It is certainly not surprising that the critical disordering is much smaller for GaAs, which has much larger quality factor. The naïve expectation of the inverse proportionality of the critical disordering and the quality factor is definitely not confirmed by our study, so $\sigma$ decreases with $N$ much slower than $1/Q$. We cannot suggest any analytical relationship between the exponential rates of changes for $\sigma$ and $Q$, because we do not have sufficient amount of data at this time.

## 5. CONCLUSION

In this paper we reported the investigation of the optical modes formed in circular one-dimensional arrays of dielectric particles. Our main interest was whether such array can possess the bound modes with the very high quality factors. We examined this question for three materials used for various optical applications including GaAs (refractive index $n_r=3.5$), TiO$_2$ ($n_r=2.7$, rutile phase) and ZnO ($n_r=1.9$). According to theoretical analysis it is possible to create modes within the circular array with the decay rate exponentially decreasing with the number of particles $N$ and we tried to find out whether this can be done using standard optical materials.

To study optical modes we employed the multisphere Mie-scattering formalism used to find eigensolutions of Maxwell equations. This approximation permits us to replace the continuous Maxwell equations by the finite set of discrete equations representing each scattering object by its multipole moments. The eigensolutions of the problem describe quasistates of the system having certain frequencies and decay rates due to the radiative decay of modes. We studied quasistates possessing the smallest decay rate or, equivalently, the highest quality factor.

Using the dipolar approximation we demonstrated that the quality factor of most bound whispering gallery modes indeed increases exponentially with the number of particles within the array for all three materials of interest. It turns out that the most bound mode is the transverse mode with the polarization perpendicular to the circle plane and characterized by the quasi-angular momentum projection $M=N/2$ to the $z$-axis perpendicular to the plane. The quality factor for this mode is very high in GaAs, where it reaches $10^{14}$ for *70* particles. It is also large for $TiO_2$ ($10^{12}$ for $N=100$), while in ZnO it does not exceed *200* even for *150* particles. In the limit of large $N$ one can specify two transverse modes and one longitudinal mode. Frequencies and quality factors for two transverse modes are quite similar, while frequencies of longitudinal modes are generally higher and their quality factors are smaller than those for transverse modes. The quality factor still increases exponentially with $N$ for longitudinal modes in GaAs and $TiO_2$, while all longitudinal whispering gallery mode in ZnO can be unbound.

The dipolar approximation turns out to be surprisingly accurate for all materials we have studied. The accuracy becomes better with larger values for the refractive index $n_r$ because the resonances are narrower at higher $n_r$ so their coupling is weaker. However, even for the lowest refractive index $n_r=1.9$ which we have studied, the correction to the resonant frequency does not exceed *2%* at large number of particles. If one defines the perturbation series formed by considering higher and higher resonances (see Fig. 4), this series appears to converge rapidly because the first non-vanishing correction is much greater then the second one. Generally, one can conclude that the dipolar approach can be used to study the low frequency optical modes in arrays of dielectric particles with reasonable refractive index $n_r>2$.

Disordering in particle sizes and placements remarkably reduces the quality factor of the system. For instance, to make a ring of *30* particles of $TiO_2$ with a quality factor exceeding $10^4$ as in the ideal sample, we need to keep the relative disordering below *1%*. The minimum acceptable disordering which does not strongly reduce the quality factor compared to the ordered system decreases exponentially with the number of particles. This is the consequence of the exponential increase of the quality factor. The rate of exponential decrease of acceptable disordering is smaller then the exponential growing rate for the quality factor.

One should notice the recent experimental work[31,32] and the related theoretical study[33] where the modes formed by coupled high frequency Mie resonances were studied. These modes definitely possesses smaller radiative decay rates at small number of particles $N$ then the low frequency modes studied in this work. On the other hand at large $N$ the modes studied in this work leaves longer because they possess larger exponential growing rate Eq. (10) increasing with the decrease of the frequency.

This work is supported by the U.S. Air Force Office of Scientific Research (Grant No. FA9550-06-1-0110). Authors acknowledge Arthur Yaghjian, Svetlana Boriskina, Alexei Yamilov, Hui Cao and Il'ya Polishchuk for useful discussions and suggestions.

## REFERENCES

1. See e. g. M. Kerker, *The scattering of light and other electromagnetic radiation*, Academic Press, New York and London, 1969.
2. E. Yablonovitch, "Inhibited spontaneous emission in solid-state physics and electronics", *Phys. Rev. Lett.* 58 (20), 2059-2062 (1987).
3. J. D. Joannopoulos, R. D. Meade, J. N. Winn, *Photonic Crystals: Molding the Flow of Light,* Princeton University Press, 1995, K. Inoue, K. Ohtaka, *Photonic Crystals: Physics, Fabrication and Applications*, Springer, 2004.
4. S. John, "Localization of light", *Physics Today*, 5, Cover story, 1991; L. I. Deych, A. A. Lisyansky, "Local polariton states in impure photonic crystals", *Phys. Rev. B* 57 (9), 5168-5176 (1998).
5. S. H. Fan, P. R. Villeneuve, J. D. Joannopolous, H. A. Haus, "Channel drop filters in photonic crystals", *Opt. Express* 3 (1), 4-11 (1998).
6. S. M. Weiss, M. Haurylau, P. M. Fauchet, "Tunable photonic bandgap structures for optical interconnects", *Opt. Mater.* 5 (2), 740-744 (2005).
7. D. A. B. Miller, "Photonic crystals: Straightening out light", *Nature Mater.* 5 (2), 83-84 (2006).
8. J. C. Knight, J. Broeng, T. A. Birks, P. H. J. Russel, "Photonic band cap guidance in optical fibers", *Science* 282 (5393), 1476-1478 (1998).
9. Z. Y. Tang, N. A. Kotov, "One-dimensional assemblies of nanoparticles: Preparation, properties, and promise", *Adv. Mater.* 17(8), 951-962 (2005).

10. H. W. Ehrespeck, H. Poehler, "A new method for obtaining maximum gain from Yagi antennas", *IEEE Trans. Antennas Propag.* AP-7, 379-386 (1959).
11. C. Kittel, *Introduction to Solid State Physics*, Wiley, New York, 1996.
12. R. A. Shore, A. D. Yaghjian, "Traveling electromagnetic waves on linear periodic arrays of lossless spheres", *Electron. Lett.* 41 (10), 578-580 (2005); L. L. Zhao, K. L. Kelly, G. C. Schatz, "The extinction spectra of silver nanoparticle arrays: Influence of array structure on plasmon resonance wavelength and width", *J. Phys. Chem. B* 107 (30), 7343-7350 (2003); S. V. Boriskina, "Spectrally engineered photonic molecules as optical sensors with enhanced sensitivity: a proposal and numerical analysis ", *J. American Opt. Soc. B* 23 (8), 1565-1573, (2006).
13. S. A. Mayer, P. G. Kik, H. A. Atwater, S. Meltzer, E. Harel, B. E. Koel, A. A. G. Requicha, "Local detection of electromagnetic energy transport below the diffraction limit in metal nanoparticle plasmon waveguides", *Nature Mater.* 2 (4), 229-232 (2003).
14. S. A. Mayer, M. L. Brongersma, P. G. Kik, S. Meltzer, A. A. G. Requicha, B. E. Koel, H. A. Atwater, "Plasmonics – A route to nanoscale optical devices", *Adv. Mater.* 15 (7-8), 562-562 (2003).
15. A. L. Burin, G. C. Schatz, H. Cao, M. A. Ratner, "High quality optical modes in low-dimensional arrays of nanoparticles. Application to random lasers", *J. American Opt. Soc. B* 21 (1), 121-131 (2004).
16. A. L. Burin, "Bound whispering gallery modes in circular arrays of dielectric spherical particles", *Phys. Rev. E* 73, 066614 (2006).
17. R. W. P. King, G. J. Fikioris, R. B. Mask, *Cylindrical Antennas and Arrays*, Cambridge University Press, Cambridge, 2005.
18. Y. L. Xu, "Electromagnetic scattering by an aggregate of spheres: far field", *Appl. Opt.* 36 (36), 9496-9508 (1997).
19. Y. L. Xu, "Scattering Mueller matrix of an ensemble of variously shaped small particles", J. American Opt. Soc. A 20 (11), 2093-2105 (2003).
20. A. Taflove, S. C. Hagness, *Computational Electrodynamics: The finite-difference time-domain method*, 3$^{rd}$ ed. Artech House Publishers, 2005.
21. D. K. Freeman, T. T. Wu, "Variational principle formulation of the two-term theory for arrays of cylindrical dipoles", *IEEE Trans. Antennas Propag.* 43 (4), 340-349 (1995).
22. L. D. Landau, E. M. Lifshitz, *Quantum Mechanics: Non-Relativistic Theory*, Pergamon Press, Oxford, New York, 1977.
23. I. S. Gradshteyn, I. M. Ryzhik, *Table of integrals, series and products*, Fifth Edition, Academic Press, London, 1994.
24. A. J. Devaney, E. Wolf, "Radiating and nonradiating classical current distributions and the fields they generate", *Phys. Rev. D* 8 (4), 1044-1047 (1973); A. Gamliel, K. Kim, A. I. Nachman, E. Wolf, "A new method for specifying nonradiating, monochromatic, scalar sources and their fields", *J. American. Opt. Soc. A* 6 (9), 1388-1393 (1989).
25. D. Margetis, G. Fikioris, J. M. Myers, T. T. Wu, "Highly directive current distributions: General theory", *Phys. Rev. E* 58 (2), 2531-2547 (1998).
26. A. Yamilov, H. Cao, "Density of resonant states and a manifestation of photonic band structure in small clusters of spherical particles", *Phys. Rev. B* 68, 085111 (2003).
27. M. I. Stockman, V. M. Shalaev, M. Moskovits, R. Botel, T. F. George, "Enhanced Raman scattering by fractal clusters: Scale-invariant theory", *Phys. Rev. B* 46 (5), 2821-02830 (1992).
28. A. L. Burin, H. Cao, M. A. Ratner, R. P. H. Chang, "Model for a random laser", *Phys. Rev. Lett.* 87, 215503 (2001).
29. J. P. Eckmann, D. Ruelle, "Ergodic theory of chaos and strange attractors", *Rev. of Mod. Phys.* 57 (3), 617-656 (1985).
30. O. P. Bruno, A. Sei, "A fast high-order solver for EM scattering from complex penetrable bodies: TE case", *IEEE Trans. Antennas Propag.* 48 (12), 1862-1864 (2000).
31. A. Yariv, "Coupled-resonator optical waveguide: a proposal and analysis", *Opt. Lett.* 24 (11), 711-713 (1999).
32. Y. Hara, T. Mukaiyama, K. Takeda, M. Kuwata-Gonokami, "Heavy photon states in photonic chains of resonantly coupled cavities with supermonodispersive microspheres", *Phys. Rev. Lett.* 94 (20), 203905 (2005).
33. L. I. Deych, A. Roslyak, "Photonic band mixing in linear chains of optically coupled microspheres", *Phys. Rev. E* 73 (3), 0366606 (2006).